\documentclass[11pt]{article}
\usepackage{amsmath}
\usepackage{amsfonts}
\usepackage{amssymb}
\usepackage{graphicx}
\usepackage{subfigure}
\usepackage{graphicx}
\usepackage{dcolumn}
\usepackage{bm}
\def\br{\biggr}
\def\bl{\biggl}

\def\be{\begin{equation}}
 \def\ee{\end{equation}}
\def\bea{\begin{eqnarray}}
\def\eea{\end{eqnarray}}
\def\f{\frac}
\def\n{\nonumber}
\def\l{\label}
\def\P{\Phi}
\def\p{\phi}
\def\o{\over}
\def\om{\omega}
\def\Om{\Omega}
\def\r{\rho}
\def\L{\Lambda}
\begin{document}
\begin{center}
\LARGE {   Ghost dark energy in $f(R)$  model of gravity}
\end{center}
\begin{center}
{\bf $^{a}$ Kh. Saaidi}\footnote{ksaaidi@uok.ac.ir},
{\bf $^{b}$A. Aghamohammadi}\footnote{ agha35484@yahoo.com},
{\bf $^{a}$B. Sabe}\footnote{ behzad.sabet@yahoo.com},
{\bf $^{c}$O. Farooq}\footnote{ {omerfarooq1900@yahoo.com}}\\
{\it$^a$Department of Physics, Faculty of Science, University of Kurdistan,  Sanandaj, Iran. }\\
{ \it $^b$Faculty of Science, Sanandaj Branch, Islamic Azad University, Sanandaj, Iran. }\\
{\it $^c$Department of Physics, Kansas State University,116 Cardwell Hall, Manhattan, KS 66506, USA.}\\

\end{center}
 \vskip 1cm
\begin{center}
{\bf{Abstract}}
\end{center}
We study a correspondence between $f(R)$ model of gravity and a phenomenological kind of dark energy (DE), which is known as QCD ghost dark energy.
Since this kind of dark energy is not stable in the context of Einsteinian theory  of gravity and Brans-Dicke model of gravity, we consider two kinds of correspondence between modified gravity
and DE. By studding the dynamical evolution of model and finding  relevant quantities such as, equation of state parameter, deceleration parameter,
dimensionless density parameter, we show that the model can describe the present Universe and also the EoS parameter can cross the phantom divide line  without needs to any kinetic energy with negative sign.
Furthermore, by  obtaining  the adiabatic squared sound speed of the model for  different cases of  interaction, we show that this model is stable.

Finally, we fit this model with supernova observational data in a non interaction case and we find the best values of parameter at $1\sigma$ confidence interval as; $f_0=0.958^{+0.07}_{-0.25}$, $\beta=-0,256^{+0.2}_{-0.1}$, and $\Om_{m_0} = 0.23^{+0.3}_{-0.15}$. These best-fit values show that   dark energy  equation of state parameter, $\om_{d_0}$, can cross the phantom divide line at the  present time.\\

\newpage
\section{Introductions}
It was indicated that our Universe is in a positive accelerating  expansion phase  \cite{1, sp, tm}, and the standard model of gravity couldn't explain this phenomena. This is a shortcoming for Einsteinian theory of gravity and several attempts have been accomplished to justify it.

 One way for explaining this  positive accelerating expansion is modified gravity. In fact some people have believed this shortcoming of Einsteinian theory of gravity is coming from the geometrical part of Hilbert-Einstein action and try to modify it by replacing a function of Ricci scalar, $f(R)$, instead of $R$ in the Hilbert-Einstein action and so-called $f(R)$ model of gravity.\footnote{There are another modified gravity formalism  which are completely different with $f(R)$ model of gravity.}
The various aspect of $f(R)$ models of gravity is investigated in  \cite{fre, so, 6,7, sc2, sn1, cs, cno, amc,  sn2, de, ak, kaa, aks, ka, f7, tm2}.

On the other hand some people looking for the reasons of this accelerating expansion in the matter part of Hilbert-Einstein action. Based on this idea, an ambiguous  component of matter so-called dark energy (DE) is introduced  for describing the positive accelerating expansion of the Universe. The simplest model of  DE is the cosmological constant which is a key ingredient in the $\Lambda$CDM model. The $\Lambda$CDM model, which is consistent with {\it nearly} all observational data, but it faces with the fine tuning problem
\cite{sw}. Also some other  DE models have been introducedand are studied in \cite{f8, pj, rr, ml}.

Recently, a new kind of  dark energy model, so called  QCD ghost dark energy (QCDGDE) model has been proposed {\cite{ 32}}. This  model, although is un-physical in the usual Minkoski space-tim, express important physical effects in dynamical space-time or space-time with non-trivial topology.
 Actually,  introducers of this dark  energy model,  claim the vacuum energy of Veneziano ghost flied in effective quantum field theory at low energy,  is related to density of dark energy which is used for explaining the positive accelerating expansion of the Universe \cite{33}. They have shown that in a curved space time, the ghost field gives rise to a vacuum energy density $H\Lambda^3_{QCD}$  of the right magnitude $\sim (10^{-3} ev)^4$, where $H$ is the Hubble parameter and $\Lambda^3$ is QCD mass scale \cite{34,341, 342, 343,344}.

  There are two groups which study  QCD ghost dark energy model. The first group which have  introduced  it have believed that  in the ghost model of dark energy, one needs not to introduce any new degree of freedom or modify gravity and it is totally embedded in standard model of gravity. The second group have  accepted the relation of density of QCDGDE as a phenomenological model of dark energy and studied the dynamical evolution of it in Einstein and Brans-Dicke model of gravity \cite{35, 35c, 35s, 351, 352, 353, 354, 355, 356}. The studies of the second  group show that the dynamical evolution of  this model is instable. This means that the QCD ghost dark energy  model has to be study further.  One can study the proposal of the first group and consider the gravitational effective field theory at low energy for some extra interaction or for some another cutoff for the size of the Universe (manifold). And the another possibility for studding this issue is based on the second group work with some extra degree of freedom or modify gravity. According to second group's view of point, we study QCDGDE model as a phenomenological model of dark energy in $f(R)$ model of gravity.

  As was mentioned earlier,   modified gravity and DE model of gravity,  are used for explaining  the positive accelerating phase of the Universe and  it is reasonable there are some correspondence between them.  Based on this idea  we separate some extra terms of modified gravity  and define correspondence relation between it and a new kind of DE. Moreover, since there are some shortcoming in these models when anybody  consider one of them alone, we studied a combination of modified gravity and dark energy model here. Therefore  we assume the matter component of the action is consist of cold dark matter (CDM)\footnote{In fact cold dark matter and ordinary matter are denoted by index "m" for simplicity.} and dark energy.
In fact, we  combine the extra terms of modified gravity with  dark energy part of matter component and equivalent it with QCD ghost dark energy.    Finally, we numerically  and analytically compute some quantities such as  DE density parameter, squared adiabatic sound speed, equation of state  and  deceleration parameter of the model.

This paper is organized as follows. In Sec.\;2, we rewriting field equations on the $f(R)$ model of  gravity and obtain the equation of motions and conservation relation for density  energy. In Sec.\;3, we make an equivalence relation between $f(R)$ model and GDE and  obtain the relevant quantities of model such as,  density parameter,  squared adiabatic speed, EOS parameter  and deceleration parameters.  At last we summarize our work and give some discussion in Sec.\;4.         

 \section{Description and general  properties of  the model}
The action of f(R) gravity with general matter is given by
\begin{equation}\label{1}
S=\int  \sqrt{-g}\; d^4x\left [\frac{f(R)}{2} + L_m(\psi, g_{\mu\nu}) \right].
\end{equation}
Where $f(R) $ is an arbitrary function of Ricci scalar, $R$, $L_m = L_m(\psi, g_{\mu\nu})$ is the matter Lagrangian, $\psi$ is the matter field, $g_{\mu\nu}$ is the metric of space-time and $g$ is the determinant of metric. Here we have  assumed  $8\pi G =1 $.  Variation of (\ref{1}) with respect of $g^{\mu\nu}$ gives
\begin{equation}\label{2}
R_{\mu\nu}f'-\frac{1}{2}fg_{\mu\nu}+\left( g_{\mu\nu}\Box-\nabla_{\mu}\nabla_{\nu}\right)f'=T^m_{\mu\nu},
\end{equation}
where prime represents  the derivative with respect to  the curvature scalar $R$, $\Box$ is the covariant  d'Alembert  operator ($\Box\equiv \nabla_{\alpha}\nabla^{\alpha}$) and $T^m_{\mu\nu}$ is the stress-energy tensor of matter which is defined by
\be\l{3}
T^m_{\mu\nu} = {-2 \o \sqrt{-g}} {\delta(\sqrt{-g} L_m )\o \delta g^{\mu\nu}}.
\ee
 The  stress-energy   tensor is covariantly conserved, this means that
 \be\l{4}
\nabla^{\mu}\bigg[ R_{\mu\nu}f'-\frac{1}{2}f g_{\mu\nu}+\left( g_{\mu\nu}\Box-\nabla_{\mu}\nabla_{\nu}\right)f'\bigg ]=\nabla^{\mu}T^m_{\mu\nu} =0.
\ee

Now we consider   a homogeneous and spatially-flat  space-time  with Friedmann-Lema$\hat{\rm i}$tre-Robertson-Walker (FlRW) line element
   \begin{equation}\label{5}
ds^2= dt^2-a^2(t)(dx^2 +dy^2+dz^2),
\end{equation}
where $a(t)$ is the scale factor.   We  will assume the Universe is filled out with    perfect fluids. Then the  stress-energy  tensor may be   given by
\begin{eqnarray}\label{6}
T_{\mu\nu}=\left(\rho_t+p_t \right)u_{\mu}u_{\nu}-g_{\mu\nu}p_t,
\end{eqnarray}
 where $\rho_t$ and $p_t$  are  the energy density  and  pressure of the fluids and $u^{\mu}=\left( 1, 0, 0, 0\right) $ is its normalized four-velocity in  co-moving  coordinates in which $u_{\mu}u^{\mu}=1$. So by substituting (\ref{5}) in to the right hand  part of (\ref{4}) and making use of (\ref{6}), one can arrive at
  \bea\l{7a}
\dot{\r}_{t}+3H(1+\om_{t})\r_{t}&=&0,
\eea
where we have used $p_t = \om_t\r_t$.
  To study the dynamics of DE model in the $f(R)$ gravity, we consider a flat  Universe with only two energy components; cold dark matter (CDM) and dark energy (DE), in which  baryon is include  in the CDM  part\footnote{In fact in this work we study a model which consist of $f(R)$ model of gravity and dark energy simultaneously.}. This means
   \be\l{8}
 \r_t = \r_m + \r_{DE},  \hspace{2cm} p_t=p_m+p_{DE},
 \ee
 so, one can decompose (\ref{7a}) as
\bea\l{9}
\r_m + 3H\r_m &=&Q,\\
\r_{DE}+3H(1+\om_{DE})\r_{DE}&=&-Q.\l{10}
\eea
where $Q$ is the direct interaction between two different components of matter.
One can show that the action (\ref{1}) is equivalent with \cite{n11}
\begin{equation}\label{11}
S=\int \frac{1}{2}\Big [f(\phi)R+\P(\phi) + 2L_m \Big ]\sqrt{-g}\;d^4x,
\end{equation}
where  $f(\phi)$ and $\Phi(\phi)$  are the  proper functions of  a scalar field $\phi$. So using  $f(R)=f(\phi)R+\P(\phi)$,   one can rewrite (\ref{2}) as
 \be\l{12}
 G_{\mu\nu} = \tilde{T}_{\mu\nu},
 \ee
 where $G_{\mu\nu}$ is the Einsteinian  tensor and
 \be\l{13}
 \tilde{T}_{\mu\nu}= {1\over f(\phi)}\Big[ T^m_{\mu\nu} + T^{\phi}_{\mu\nu}\Big],
 \ee
 and
 \be\l{14}
 T^{\phi}_{\mu\nu}= \bigg [ {1\over 2}g_{\mu\nu}\Phi(\phi) + (\nabla_{\mu}\nabla_{\nu}- g_{\mu\nu} \Box)f(\phi) \bigg].
 \ee

 It is well known that  $\nabla^{\mu}G_{\mu\nu} =0$, them  from (\ref{12}) we have $\nabla^{\mu}\tilde{T}_{\mu\nu} =0$. Since,     $\nabla^{\mu}T^m_{\mu\nu} =0$,  so we can obtain
 \be\l{15}
 \nabla^{\mu}T^{\p}_{\mu\nu}=\bigg[T^{m}_{\mu\nu} +T^{\p}_{\mu\nu}\bigg]\nabla^{\mu}(\ln f),
 \ee
 The $tt$ component of (\ref{15}) and Eqs. (\ref{9}) and (\ref{10}) gives
 \bea \l{16}
  \dot{\rho}_{m} &+& 3H\rho_{m} =Q\\
  \dot{\rho}_{DE} &+& 3H(\rho_{\phi} + {p}_{DE}) =-Q,\l{17},\\
 \dot{\rho}_{\phi} &+& 3H(\rho_{\phi} + \check{p}_{\phi}) =0,\l{18}
 \eea
 where
 \be\l{19}
 \check{p}_{\phi} = p_{\p} -{\dot{f} \o 3Hf}\big [\rho_m + \r_{\p}\big ].
 \ee
As already mentioned in the Introduction, we suppose $\r_{\p}$ has the same as dark energy role in  evolution of the  Universe. So we can combine it with the dark energy part which is coming from the matter part (the right hand side) of Einstein equation, (\ref{2}). This means that we can rewrite (\ref{16} ), (\ref{17}) and (\ref{18}) as follows
  \bea \l{20}
  \dot{\rho}_{m} &+& 3H\rho_{m} =Q\\
  \dot{\rho}_{\L} &+& 3H(\rho_{\L} + {p}_{\L}) =-Q,\l{21},
  \eea
 where
 \be\l{22}
 \r_{\L} = \r_{\p} + \r_{DE},  \hspace{2cm} p_{\L}=\check{p}_{\p}+p_{DE},
 \ee
  The $tt$ component of the  gravitational equations (\ref{12}), for the metric (\ref{5}), can be simplified to
   \begin{eqnarray}\label{23}
H^2=\frac{1 }{3f(\p) }\left( \rho_{t}+\rho_{\p}\right),
\end{eqnarray}
where
\begin{equation}\label{24}
\rho_{\phi}= \frac{1}{2}{\P(\phi)-3H\dot{f}(\p)  }.
\end{equation}
Using (\ref{23}) we arrive at
 \begin{eqnarray}\label{25}
\dot{H}=-{1\o 2f(\p)}\bigg[ \rho_{t}+\r_{\p} +p_{DE}+ p_{\p}\bigg ].
\end{eqnarray}
\section{Dynamics of ghost dark energy}
 In this work we want to consider a correspondence between $f(R)$ model of gravity and ghost dark energy.
In the   ghost model of dark energy, the energy density of dark energy is given by $\rho_{d}=\alpha H$ where $\alpha$ is a constant with dimension $[energy]^3$ roughly of order of $\Lambda^3_{QCD}$, where $\Lambda_{QCD}\sim 100$ MeV
is $QCD$ mass scale \cite{32}.\footnote{ Note that the relevant  physical quantities, $\om_d$, $q$, $\Om_{ed}$, $c^2_s$,  are independent of $\alpha$.} For this correspondences we have two choices as follows
\begin{itemize}
\item  We define a correspondence relation as \\
\be\l{26}
\r_{d} = \r_{\p} + \r_{DE}, \hspace{2cm}p_{d} = p_{DE}+\check{p}_{\p}.
\ee
Here $\r_{d}$ and $p_{d}$ are density energy and pressure of ghost dark energy respectively. In  this case  the  relation of conservation  for matter and dark energy is the same  as (\ref{20}) and (\ref{21}).

 \item Another possibility for a correspondence relation is \\
\be\l{27}
\r_{d} = \r_{\p} + \r_{DE}, \hspace{2cm}p_{d} = p_{DE}+{p}_{\p}.
\ee
In  this case   the  relation of conservation for matter and dark energy is as follows
 \bea \l{28}
  \dot{\rho}_{m} &+& 3H\rho_{m} =Q,\\
 \dot{\rho}_{d} &+& 3H(\rho_{d} + {p}_{d}) ={\dot{f} \o f}\big [ \r_m + \r_d\big]-Q,\l{29}
 \eea
 in this case we have two different interactions. Q interaction is a direct interaction between CDM and DE and the other term is coming from the interaction between matter and geometry.

\end{itemize}

\subsection{Dynamics of ghost dark energy for $p_{d} = p_{\L}$}
Some observational data such as, observational of  the galaxy cluster Abell A586,   supports the interaction between DE and CDM \cite{23s}. Therefore in this section we  study the direct interaction between DE and
CDM and study the  dynamical evolution  of the  model. So  in   (\ref{20}) and (\ref{21}), we assume $Q\neq 0$. Then  by $\r_{d} =\om_{d}p_{d}$ we have
\begin{eqnarray}\label{30}
  \dot{\rho}_m+ 3H\rho_m &=& Q \\
  \dot{\rho}_{d}+3(1+\omega_{d})H\rho_{d} &=& -Q.\l{31}
\end{eqnarray}
 One can rewrite (\ref{23}), (\ref{30}) and (\ref{31}),  in terms of  dimensionless quantities as
\begin{eqnarray}
  \Omega_{em} +\Omega_{ed}&=& 1, \label{32} \\
 \dot{\Omega}_{em}+{2\dot{H} \o H}\Om_{em}+(3 +\beta)H\Omega_{em} &=&   \frac{ Q}{3H^2 f}, \label{33}\\
 \dot{\Omega}_{ed}+{2\dot{H} \o H}\Om_{ed}+(3 +\beta + 3\om_D)H\Omega_{ed} &=& -  \frac{ Q}{3H^2 f},\label{34}
\end{eqnarray}
where $\Omega_{em}=\rho_m/3H^2f(\phi),\, \Omega_{ed}=\rho_{d}/3H^2f(\phi)$ are   the dimensionless energy density
of CDM and DE respectively and $ \dot{f} = \beta f H$.  Using (\ref{32}) and (\ref{33}) we obtain
\be
-\dot{\Omega}_{ed}+{2\dot{H} \o H}(1-\Om_{ed})+(3 +\beta )H(1-\Omega_{em} )=  \frac{ Q}{3H^2 f},\label{35}
\ee
and substituting (\ref{34}) into (\ref{35}) we have
\be\l{36}
{2\dot{H} \o H} +(3+\beta)H + 3H\om_d \Om_{ed} = 0.
\ee
As well, based on  the linear relation between $\rho_{d}$ and H, we have
 \begin{eqnarray}\label{37}
H\Omega_{ed}f= {\rm constant},
  \end{eqnarray}
   Expressing (\ref{35}) in terms of efolding-number $x\equiv \ln a$,  and making use  (\ref{37}) we have
   \be\l{38}
   -\Omega'_{ed}{2 - \Omega_{ed} \over \Omega_{ed}}={Q \o 3fH^3}+{(\beta-3)(1 - \Omega_{ed})}.
   \ee
   Here, the equation of state parameter of the  model   versus $\Omega_{ed}$ is
   \be\l{39}
   \om_d = -{1 \o 3}\big[{ 3+ 2\Om_Q -\beta \o   2 - \Om_{ed}}\big],
   \ee
  where $\Om_Q = Q/(3f\Om_{ed}H^3)$ and deceleration parameter $q=-1-\dot{H}/H^2$ is
    \be\l{40}
   q = {\beta +1-(\Om_Q+2)\Om_{ed} \o(  2 - \Om_{ed})}.
   \ee
   Also    we can obtain the squared adiabatic sound speed of  our model as
   \bea\l{41}
   c_s^2 &=& {dp_{d} \o d\rho_{d}} = {\Bigg [}1-\bigg\{ {(3-\beta) - (3+\Om_Q -\beta)\Om_{ed} \o (3+\beta) - (3 +\Om_q)\Om_{ed}}\bigg\} \Om_{ed}{d \over d\Om_{ed}}{\Bigg ]}\omega_{d}.
   \eea
   For getting better insight we consider the interacting GDE for three different forms of $Q$ below.
 \subsubsection{$Q = 3b^2H\rho_d$}

  In this case $\Om_Q = 3b^2 $ and Eq.  (\ref{38})  reduces to\\
 \be\l{42}
   \Omega'_{ed}{2 - \Omega_{ed} \over \Omega_{ed}}=(\beta-3 -3b^2)\big[\Omega_{ed}+K\big],
   \ee
  where
  $$K= {3-\beta \o (\beta - 3-3b^2)},$$ and by integrating of it we find
  \be\l{43}
  2\ln(\Om_{ed}) -(2+K)\ln(|\Om_{ed} +K|) = (3-\beta )x + C_1,
  \ee
  where $C_1$ is the constant of integration and it is given by  $$C_1 = 2\ln(\Om_{ed_0}) - (2+K)\ln(|\Om_{ed_0} +K|) $$ and $\Om_{ed_0}$ is the effective fraction of dark energy at the present  time.

  Also Eqs.\;(\ref{39}), (\ref{40}) and (\ref{41}) reduce to
 \be\l{44}
   \om_d = -{1 \o   2 - \Om_{ed}} -{  6b^2 -\beta \o  3( 2 - \Om_{ed})},
   \ee

 \be\l{45}
   q = {\beta +1 \o(  2 - \Om_d)}-{(3b^2+2)\Om_{ed} \o(  2 - \Om_d)},
   \ee
 \bea\l{46}
   c_s^2 &=&  {\Bigg [}\bigg\{ {(3-\beta) - (3+3b^2 -\beta)\Om_{ed} \o (3+\beta) - (3 +3b^2)\Om_{ed}}\bigg\}{ \Om_{ed}\o (2-\Om_{ed})}- 1 {\Bigg ]}{A_1 \o (2-\Om_{ed})},
   \eea
  where $A_1=(3+6b^2-\beta)/3$. For getting better insight, we compute $\om_d$, $q$ and $c^2_s$ for $\beta=-0.2$, $b=0.1$ and $\Om_{ed }=0.8$ and obtain $\om_d = -1.15$, $q=-0.99$ and $c^2_s=0.06$. It is seen that for this especial choice, the equation of state cross the phantom divide line ($\om_d=-1)$ and the adiabatic squared sound speed is positive. This means that this model can describe the positive accelerating expansion of the Universe and also it is stable. We solve numerically (\ref{43}) and plot it in Fig.\;1.  Fig.\;1a is the effective fraction of dark energy, $\Om_{ed}$, versus efolding number, $x= \ln(a)$, and shows   that at the early time $\Om_{ed} = 0$ and the late time  is saturated to 1.  Also at the present time  $\Om_{ed}= 0.8$. Fig.\;1b is the adiabatic squared sound speed versus $\Om_{ed}$. It is obviously seen that at the present time namely for $\Om_{ed}=0.8$, $c^2_s >0$, and then for this kind of interaction the model  is stable.

\begin{figure}[t]\label{7}
\begin{minipage}[b]{1\textwidth}
\subfigure[\label{fig1a} ]{ \includegraphics[width=.45\textwidth]%
{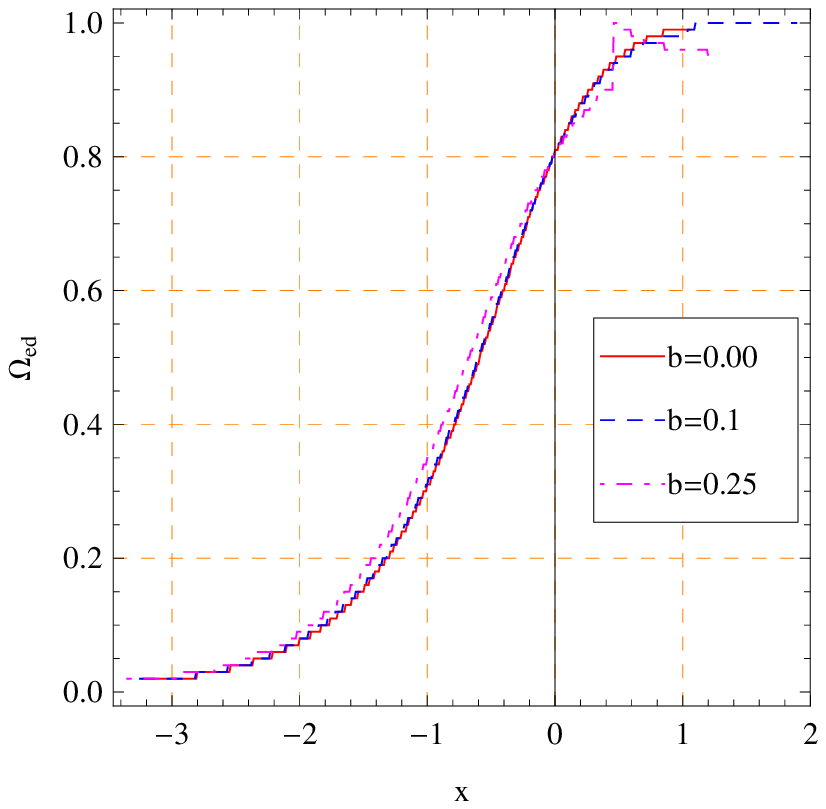}} \hspace{1cm}
\subfigure[\label{fig1b} ]{ \includegraphics[width=.45\textwidth]%
{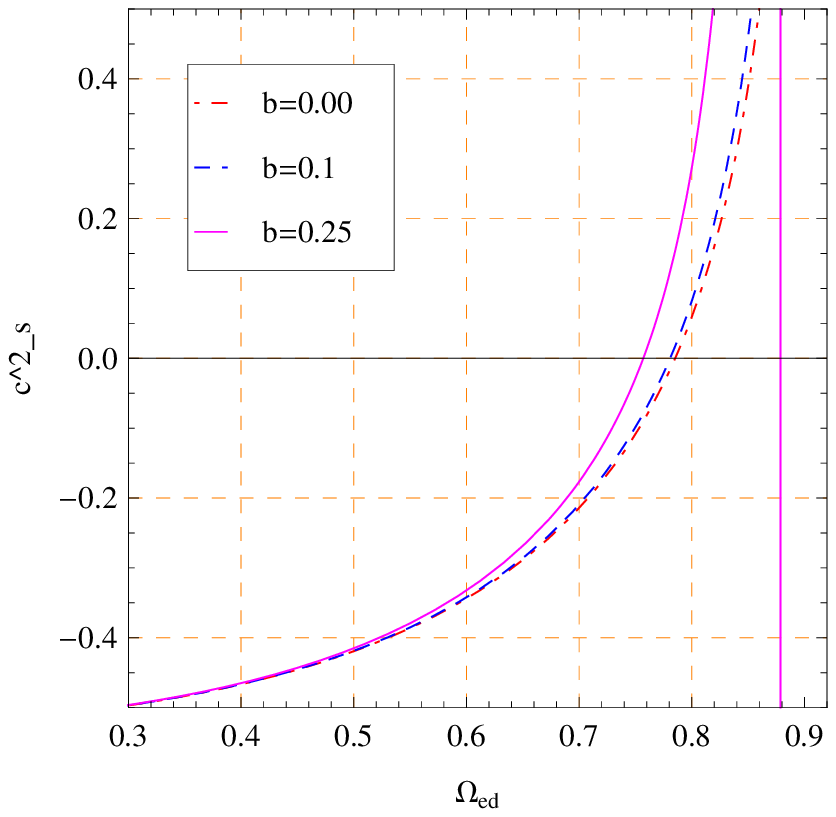}}
\end{minipage}
\caption{(a): This subfigure shows $\Om_{ed}$ versus efolding number, $x=\ln(a)$. (b): This subfigure shows $c^2_s$ versus $\Om_{ed}$. We have taken $\beta = -0.2$, $p_d=p_{\Lambda}$  and $Q= 3b^2H\rho_d$.  }
\end{figure}
\begin{figure}[t]\label{8}
\begin{minipage}[b]{1\textwidth}
\subfigure[\label{fig1a} ]{ \includegraphics[width=.45\textwidth]%
{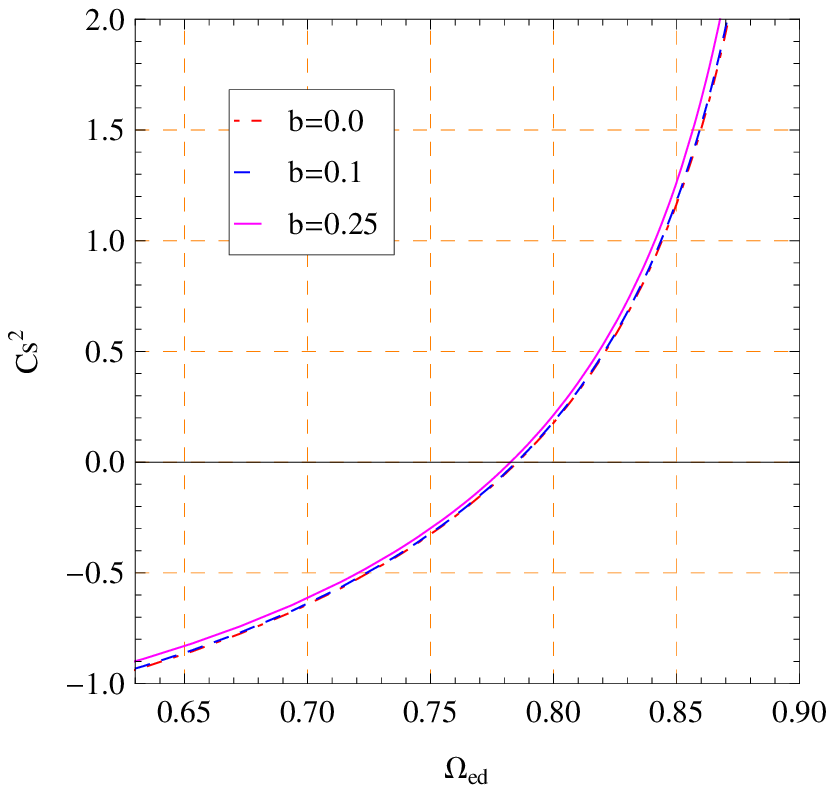}} \hspace{1cm}
\subfigure[\label{fig1b} ]{ \includegraphics[width=.45\textwidth]%
{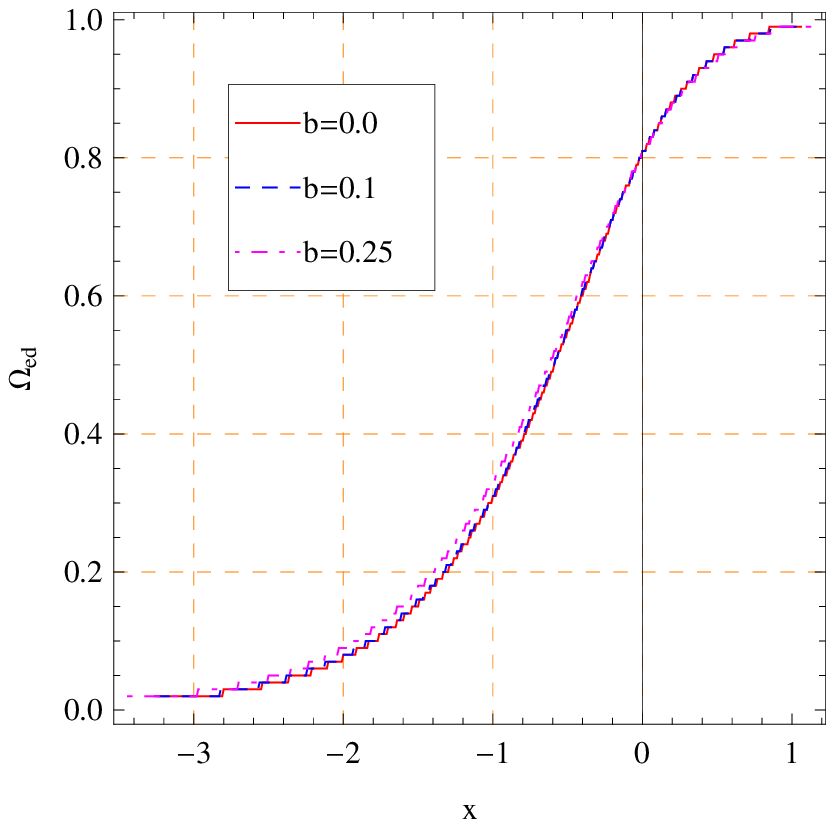}}
\end{minipage}
\caption{(a): This subfigure shows $\Om_{ed}$ versus efolding number, $x=\ln(a)$. (b): This subfigure shows $c^2_s$ versus $\Om_{ed}$. We have taken $\beta = -0.2$, $p_d=p_{\Lambda}$ and $Q= 3b^2H\rho_m$.   }
\end{figure}
\subsubsection{ $Q=3b^2H\rho_m$}
In this case $\Om_Q = 3b^2(1 -\Om_{ed} )/\Om_{ed} $ and equation (\ref{38}) reduce to\\
\be\l{47}
\Om'_{ed} {2-\Om_{ed} \o \Om_{ed}} = \xi (1- \Om_{ed}-1),
\ee
where $$\xi = {3-\beta-3b^2  }.$$
By solving (\ref{47}) we have
\be\l{48}
{\Om^2_{ed} \o |1-\Om_{ed}|} = c_2e^{\xi x},
\ee
where $c_2$ is the constant of integration and it is given by
$$c_2 = {\Om^2_{ed_0} \o |1-\Om_{ed_0}|}, $$
where $\Om_{ed_0}$ is the value of effective density parameter  of dark energy at the present time. And  Eqs. (\ref{39}) and (\ref{40})  become
 \be\l{49}
   \om_d = -{ \xi\o 3( 2 - \Om_{ed})}- { 2b^2 \o  \Om_{ed}( 2 - \Om_{ed})},
   \ee
 \be\l{50}
   q = {\beta +1-(2-3b^2)\Om_{ed} \o(  2 - \Om_{ed})},
   \ee
and the adiabatic square sound speed Eq. (\ref{41}),  is
\bea\l{51}
   c_s^2= {\Bigg [}\bigg\{ {\xi(1-\Om_{ed})\big[\xi\Om^2_{ed} +4b^2(1-\Om_{ed})\big] \o \big[(3+\beta - 3b^2) - (3 -3b^2)\Om_{ed}\big](2-\Om_{ed})}\bigg\}- (\xi \Om_{ed} + 2b^2) {\Bigg ]}{1 \o 3\Om_{ed}(2-\Om_{ed})}.
   \eea
  We solved numerically Eq.\;(\ref{48}) and plot it in Fig.\;2.  Fig.\;2a indicates the effective fraction of dark energy, $\Om_{ed}$, versus efolding number, $x= \ln(a)$ for $Q=3b^2H\rho_m$. This subfigure  shows   that the behavior of effective dimensionless parameter of dark energy is the same as  $\Om_{ed} $  for $Q=3b^2H\rho_d$.  Fig.\;2b indicate  the adiabatic squared sound speed versus $\Om_{ed}$ in the case of $Q=3b^2H\rho_m$. It is obviously seen that at the present time namely  $c^2_s >0$, and then for this kind of interaction the model  is stable too.

\subsubsection{ $Q=3b^2H\rho_t$}
In this case $\Om_Q = 3b^2/\Om_{ed} $ and equation(\ref{37})  reduces to
\be\l{52}
\Om'_{ed} {2-\Om_{ed} \o \Om_{ed}} = (3-\beta) (K- \Om_{ed}),
\ee
where $$K = {3-\beta-3b^2 \o 3-\beta }.$$
By solving (\ref{50}) we obtain
\be\l{53}
2\ln\Om_{ed} + (K-2) \ln(|\Om_{ed} - K|) = (3-\beta) K x + c,
\ee
where $$c =2\ln\Om_{0} + (K-2) \ln(|\Om_{0} - K|),  $$  is  the constant of integration.
 \be\l{54}
   \om_d = -{1 \o 3}\big[{( 3 -\beta)\Om_{ed}+ 3b^2 \o  \Om_{ed}( 2 - \Om_{ed})}\big],
   \ee
 \be\l{55}
   q = {\beta +1-3b^2 -2\Om_{ed} \o(  2 - \Om_d)},
   \ee
   and
\bea\l{56}
   c_s^2&=& {\Bigg [}\bigg\{ {(3-\beta-3b^2)-(3-\beta)\Om_{ed} \o (3+\beta - 3b^2) - 3\Om_{ed}}\bigg\}{(3-\beta)\Om^2_{ed} -6b^2(1-\Om_{ed})\ \o(2-\Om_{ed})}\n\\&&- \big[(3-\beta)\Om_{ed} + 3b^2\big] {\Bigg ]}{1 \o 3\Om_{ed}(2-\Om_{ed})},
   \eea
\begin{figure}[t]\label{7}
\begin{minipage}[b]{1\textwidth}
\subfigure[\label{fig1a} ]{ \includegraphics[width=.45\textwidth]%
{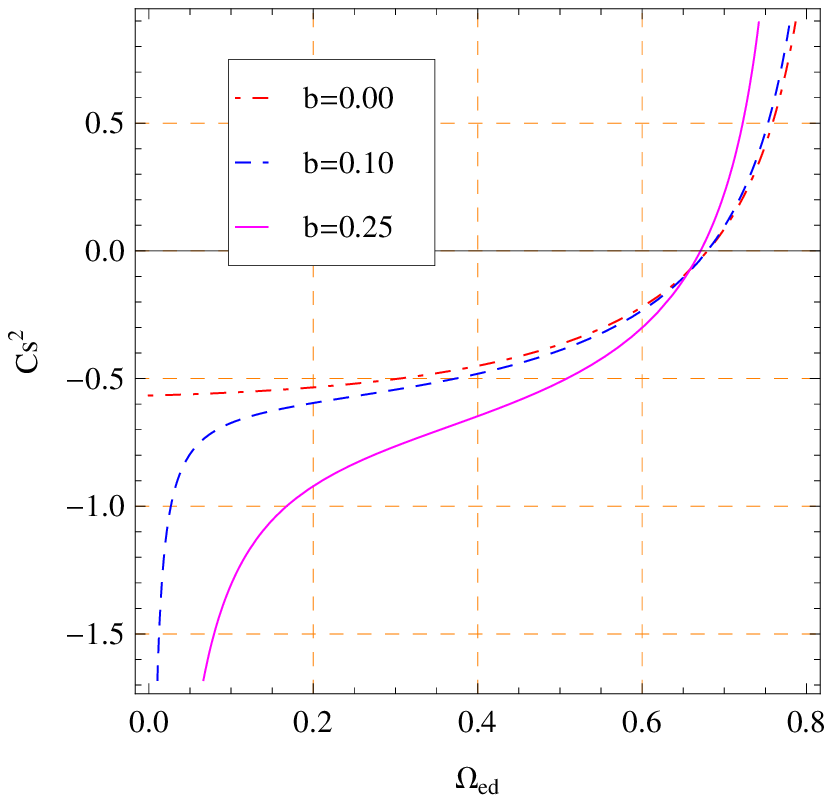}} \hspace{1cm}
\subfigure[\label{fig1b} ]{ \includegraphics[width=.45\textwidth]%
{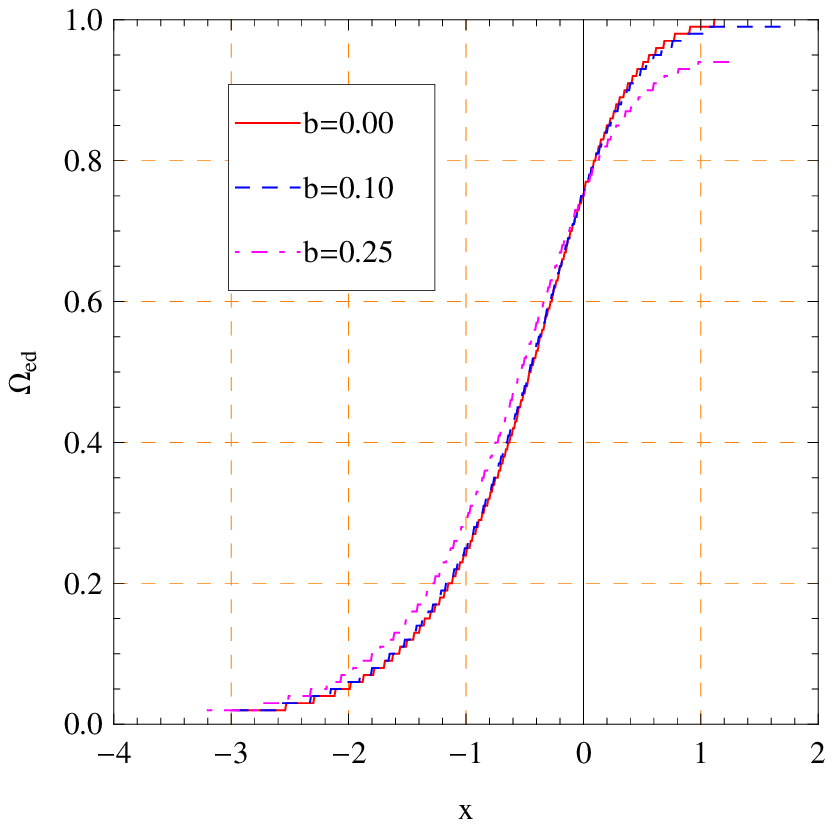}}
\end{minipage}
\caption{(a): This subfigure shows $\Om_{ed}$ versus efolding number, $x=\ln(a)$. (b): This subfigure shows $c^2_s$ versus $\Om_{ed}$. We have taken $\beta = -0.2$, $p_d=p_{\Lambda}$ and $Q= 3b^2H\rho_t$. }
\end{figure}
\begin{figure}[t]\label{8}
\begin{minipage}[b]{1\textwidth}
\subfigure[\label{fig1a} ]{ \includegraphics[width=.45\textwidth]%
{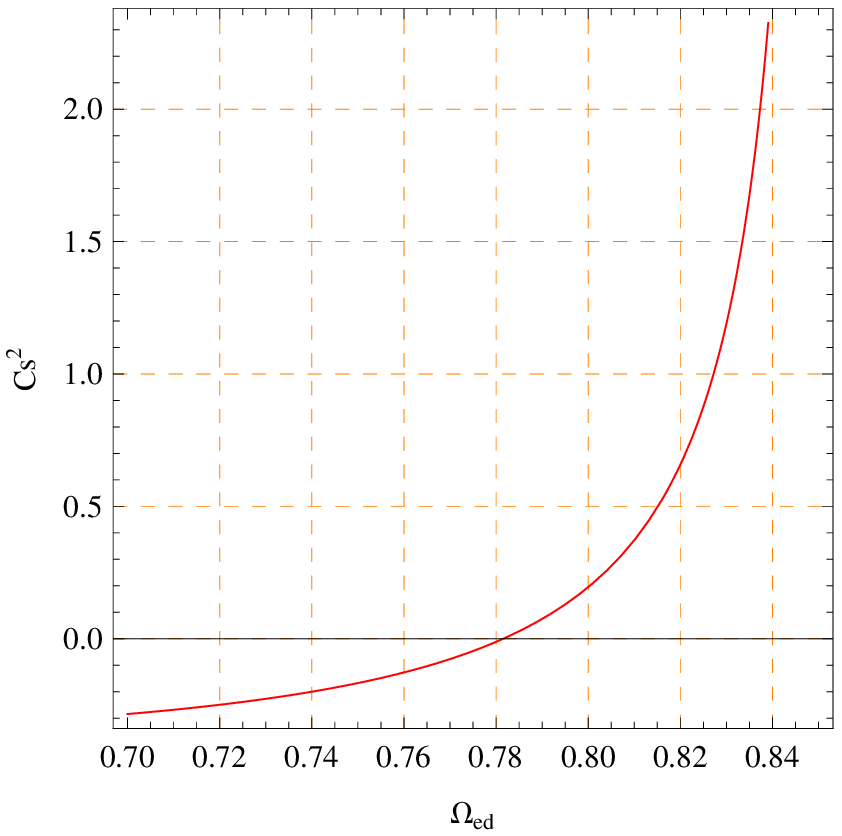}} \hspace{1cm}
\subfigure[\label{fig1b} ]{ \includegraphics[width=.45\textwidth]%
{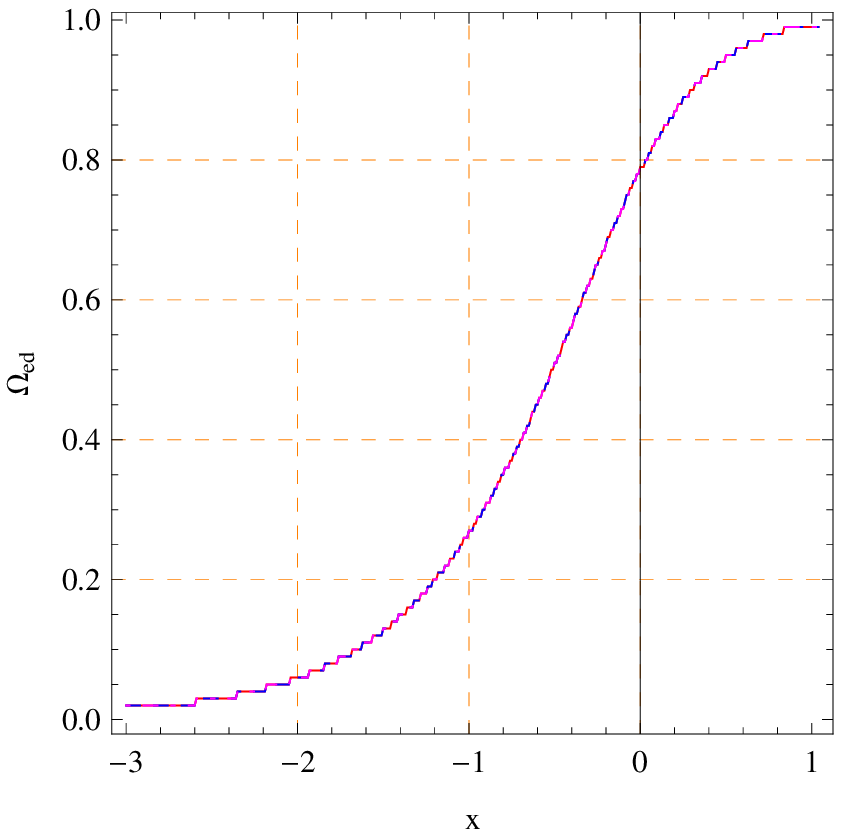}}
\end{minipage}
\caption{(a):  This subfigure shows $\Om_{ed}$ versus efolding number, $x=\ln(a)$. (b): This subfigure shows $c^2_s$ versus $\Om_{ed}$. We have taken $\beta = -0.2$, $p_d=p_{DE}+p_{\p}$ and  $Q= 0$.  }
\end{figure}

 We solved numerically Eq.\;(\ref{53}) and plot it in Fig.\;3.  Fig.\;3a indicates the effective fraction of dark energy, $\Om_{ed}$, versus efolding number, $x= \ln(a)$ for $Q=3b^2H\rho_t$. This subfigure  shows   that the behavior of effective dimensionless parameter of dark energy is the same as  $\Om_{ed} $ in the two pervious subsection.  Fig.\;2b indicate  the adiabatic squared sound speed versus $\Om_{ed}$ in the case of $Q=3b^2H\rho_t$. This subfigure shows that    at the present time namely  $c^2_s >0$, and then for this kind of interaction the model  is stable too.
\subsection{Dynamics of Ghost dark energy for $p_d = {p}_{\p}+ p_{DE}$}
In this subsection dark energy is assumed to be without direct  interaction with the matter part namely $Q=0$ . So according to (\ref{29}) dark energy  interaction may be  with  geometry. However the conservation relation of energy is as (\ref{28}) and (\ref{29}), so
 \bea \l{57}
  \dot{\r}_m &+& 3H\rho_m =0,\\
 \dot{\r}_d &+& 3H(\r_d + p_d) =\beta H \big [ \r_m + \r_d \big ].\l{58}
 \end{eqnarray}
In this case  Eqs. (\ref{33}) and (\ref{34}) can be rewrite as
\bea \l{59}
  \dot{\Om}_{em}  +\f{2\dot{H}}{H}\Om_{em}+(3+\beta)H\Om_{em}&=&0,\\
 \dot{\Om}_{ed}-\beta H r\Om_{ed}+3H\Om_{ed}+3\om_d H\Om_{ed}+2\f{\dot{H}}{H}\Om_{ed}&=&0.\l{60}
 \eea
Using Eqs.\;(\ref{32}) and (\ref{59}) one can obtain
\begin{eqnarray}\l{61}
-\dot{\Om}_{ed}+\f{2\dot{H}}{H}(1-\Om_{ed})+(3+\beta)H(1-Â\Om_{ed})=0.
\end{eqnarray}
By combining Eq.\;(\ref{60}) with Eq.\;(\ref{61}),  we have
\bea\l{62}
\f{2\dot{H}}{H}+3\om_{d}\Om_{ed}H+(3+\beta)H -(1+r)\beta H\Om_{ed}=0.
\eea
 Hence one can achieve in an expression for $\Om_{ed}$ in terms of efolding-number $x\equiv \ln a,$ as
\bea\l{63}
-\Om'_{ed}{\br(}\f{2-\Om_{ed}}{\Om_{ed}}{\br)}+{\br(}1-\Om_{ed}{\bl)}{\br[}3-\beta{\bl]}=0.
\eea
By solving (\ref{63}) we have
\be\l{64}
{\Om^2_{ed} \o |1-\Om_{ed}|} = ce^{\xi x},
\ee
where $\xi = 3-\beta$ and $c$ is the constant of integration.
  Eventually, in this case, the expression of the EoS,  $\om_d$,  of DE versus $\Om_{ed}$  is specified as follows
  \be\l{65}
   \om_{d}=-\f{1}{3}{\br[}\f{3-\beta}{2-\Om_{ed}}-\beta(1+ r){\bl]},
   \ee
   and deceleration parameter is as
   \be\l{66}
      q=\f{1+\beta-2\Om_{ed}}{2-\Om_{ed}}
   \ee

At last, we can find  the squared adiabatic sound speed of the  model as follows
\be\l{67}
  c^2_s= {1\over 3}\Bigg[ \bigg\{ {(3-\beta)(1-\Om_{ed}) \o (3+\beta) -3\Om_{ed}}\bigg\} {\Om_{ed} \o (2-\Om_{ed})}-\beta(1+r)- {3-\beta\o (2-\Om_{ed})} \Bigg],
  \ee
The numerical results of Eq.\'(\ref{64}) and Eq.\;(\ref{67}) is plotted in Fig.\;4. In this case we don't have any direct interaction between CDM and DE.
This figure shows that the effective dimensionless energy density of dark energy is started from 0 at the early time and increased to asymptotic value, $1$ at late time. Fig.\;1b indicates the adiabatic squared sound speed versus $\Om_{ed}$ and shows that at the present time, $\Om_{ed}= 0.8$, $c^2_s$ is positive. So as was expected, the model is stable in this case too.
\section{Data fitting}
{\bf In this Section we use the 557 Uion $II$ sample dataset of SnIa[], to find the best-fit forms of our model. We consider a model which includes dark energy, cold dark matter (CDM), baryon (B) and radiation (R) in a flat FLRW universe. For simplicity we write the energy density of baryon and cold dark matter  together as $\Om_m = \Om_{CDM} + \Om_{B}$. Also, since at present time the dimensionless energy density of radiation is very small compared to $\Om_m$, we neglect $\Om_{R}$. Therefore in this case the Friedmann equation is
\be\l{68}
H^2 = {1 \o 3 f(t)} ( \r_d+\r_m).
\ee
Using Eqs.\;(\ref{57}) and $\r_d =\alpha H$, one can rewrite (\ref{68}) as
 \be\l{69}
H^2 = {1 \o 3 f(t)} (\alpha H +\r_{m_0} a^{-3}).
\ee
We can use  $\dot{f} =\beta f H$ ($f= f_0 a^{\beta}$), and Eq.\:(\ref{69}) can be rewritten as
\bea\l{70}
H(f_0, \beta, \Om_{m_0})& =& H_0 \Bigg[ {1\o 2} \Big( 1 - {\Om_{m_0}(1+z)^{\beta}\o f_0}\Big)\\
 &+& \sqrt{{1\o 4} \Big( 1 - {\Om_{m_0}(1+z)^{\beta}\o f_0}\Big)^2 +{\Om_{m_0}(1+z)^{3+\beta}\o f_0}} \Bigg]\n,
\eea
In this model we have three free parameters $f_0$, $\beta$, and $\Om_{m_0}$.
To obtain the best-fit for free parameters we have to compare the theoretical  distance modulus, $\mu_{th}$,  with  observed  $\mu_{ob}$ of supernova. The distance modulus is defined by
\be\l{71}
\mu_{th} = 5{\rm log_{10}} \Big[ D_L(z; f_0, \beta, \Om_{m_0}) \Big]+ \mu_0,
\ee
where $\mu_0 = 43.3$, and $ D_L(z; f_0, \beta, \Om_{m_0})$ is given by
\be\l{72}
 D_L(z; f_0, \beta, \Om_{m_0})= (1+z)\int_0^z {H_0 \o  H(x; f_0, \beta, \Om_{m_0})}dx,
 \ee
 For comparing $\mu_{th}$ with $\mu_{ob}$ we need to obtain $\chi^2_{sn}$ which is defined by
 \be\l{73}
 \chi^2_{sn}( f_0, \beta, \Om_{m_0}) = \sum_{i=1}^{557} {\big[ \mu_{th}(z_i) - \mu_{ob}(z_i)\big]^2 \o \sigma^2_i}.
 \ee
 A minimization of this expression leads to
 \be\l{74}
 \chi^2_{sn_{min}}( f_0=0.958, \beta=-0.26, \Om_{m_0})=0.23) = 543.583,
 \ee
where implies $\chi^2_{sn}/{\rm dof} = \chi^2_{sn_{min}}/{\rm dof} = 0.981$ (dof = 554). This show that this model is clearly consistent with the data since  $\chi^2/{\rm dof} =1$.

The $1\sigma$ errors on the predicted value of free parameter,  $a$ is found by solving the following equation
\be\l{75}
\chi^2(a_{1\sigma}, b_{bf}, c_{bf}) -\c
hi^2_{min}(a_{bs}, b_{bf}, c_{bf}) =1,
\ee
where $bs\equiv best-fit value$ and  $x_{bs}$'s are the values for free parameters which minimize $\chi^2$. Using Eq.\;(\ref{75}), we found the best-fit values  and errors for free parameters at $1\sigma$ have been reported in table.I\\

\begin{table}[h]
\centering
\begin{tabular}{|c|c|c|c|}
  \hline
  parameter & $f_0$ & $\beta$ & $\Om_{m_0}$ \\
  \hline \hline
  best-fit$^{+1\sigma}_{-1\sigma}$ & 0.958$^{+0.07}_{-0.25}$ & -0.256$^{+0.2}_{-0.1}$ & 0.23$^{+0.3}_{-0.15}$ \\
  \hline
 \end{tabular}
\caption{ The best-fit values with $1\sigma$ errors for $f_0$, $\beta$, and $\Om_{m_0}$ in the $f(R)$ GDE model.}
\end{table}

\begin{figure}[ht]\label{0}
\centerline{ \includegraphics[width=7cm] {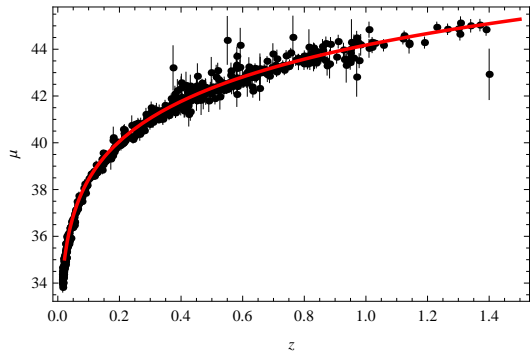}}
\caption{   The observed distance modulus of supernova (points) and the theoretical predicted distance modulus (red-solid line) in the context of $f(R)$ GDE model. }
\end{figure}
Note that according to the best-fit values of parameters the present  dark energy equation of state, $\om_{d_0}= -0.987$.   In figure 5 we show a comparison between theoretical distance modulus and observed distance modulus of supernova data. The red-solid line indicates the theoretical value of distance modulus, $\mu_{th}$, for the best value of free parameters (table 1). This figure show a reasonable result.}

\section{Conclusion}
 We  investigated two kinds  of  equivalence between $f(R)$ model of gravity and an phenomenological kind of dark energy so- called ghost dark energy whose energy density is proportional to Hubble parameter.  We studied the interacting and non interacting case of model in a flat FLRW background. For getting better results,   we consider three kinds of interaction between matter and dark energy.  The obtained results for  equation of state  and deceleration parameter, show that the model can describe the accelerating expansion phase of the Universe  and also $\om_d$ can cross the line $\om_d=-1$ from quintessence to phantom model for all cases of model with/without interaction. Also by obtaining  the numerical results for effective dimensionless energy density of dark energy, $\Om_{ed}$,  we find that $\Om_{ed}$ is started from 0 at the early time and saturated to its  asymptotic value, $1$,  at late time. We further studied the dynamical evolution of the model by considering the adiabatic squared sound speed. This quantity is obtained  for all cases of interaction and without interaction. Our investigation show that the adiabatic squared sound speed can be positive and then the model can be stable by a suitable choice of parameters.

{\bf  Also we fitted this model with supernova observational data in a non interaction case. In this case the best values of parameter  of the model are $f_0=0.958^{+0.07}_{-0.25}$, $\beta=-0,256^{+0.2}_{-0.1}$, and $\Om_{m_0} = 0.23^{+0.3}_{-0.15}$. These best-fit values show that   the present dark energy  equation of state parameter, $\om_{d_0}$, can cross the phantom divide line.}

 At last, we conclude that the dynamical behavior of the model for two different correspondence relations,  also for all  interactions/non interaction cases  have the same behavior.  This model is stable and can describe the present Universe.

\end{document}